\documentstyle[12pt]{article}
\newcommand{\Ir}{Z\!\!\!Z}
\newcommand{\idty}{{\leavevmode{\rm 1\mkern -5.4mu I}}}
\newcommand{\Ibb}[1]{ {\rm I\ifmmode\mkern -3.6mu\else\kern -.2em\fi#1}}
\newcommand{\ibb}[1]{\leavevmode\hbox{\kern.3em\vrule
     height 1.2ex depth -.3ex width .2pt\kern-.3em\rm#1}}
\newcommand{\Cx}{{\ibb C}}
\newcommand{\Rl}{{\Ibb R}}
\normalbaselines

\begin{document}
\begin{tabbing}
\hspace*{12cm}\= GOET-TP 98/93 \\
              \> December 1993
\end{tabbing}
\count0 = 1

\begin{center}
\vspace*{1.0cm}
{\LARGE \bf Differential Calculus and   \\
\vskip.3cm
     Discrete Structures}
\vskip 1.cm
{\large \bf Aristophanes Dimakis}
\vskip 0.5 cm
Department of Mathematics, University of Crete, GR-71409 Iraklion
\vskip 0.5 cm
and
\vskip 0.5 cm
{\large \bf  Folkert M\"uller-Hoissen}
\vskip 0.5 cm
Institut f\"ur Theoretische Physik, D-37073
            G\"ottingen, Germany
\end{center}

\vspace{1 cm}

\begin{abstract}
There is a deformation of the ordinary differential calculus which
leads from the continuum to a lattice (and induces a corresponding
deformation of physical theories). We recall some of its features
and relate it to a general framework of differential calculus on
discrete sets. This framework generalizes the usual (lattice)
discretization.
\end{abstract}

\vskip1cm

\begin{center}
\begin{minipage}[t]{10cm}
To appear in the proceedings of the International Symposium
on ``Generalized Symmetries in Physics'', ASI Clausthal, July 1993.
\end{minipage}
\end{center}

\newpage

\renewcommand{\theequation} {\arabic{section}.\arabic{equation}}

\section{Introduction}
\setcounter{equation}{0}
In the context of `noncommutative geometry' \cite{Conn86,Coqu89} the
following structure -- which generalizes the notion of differential
forms (on a manifold) -- plays a crucial role.
A {\em differential calculus} for an associative
algebra $\cal A$ (over $\Rl$ or $\Cx$) is a $\Ir$-graded associative
algebra $\Lambda({\cal A})=\bigoplus_{r=0}^\infty \Lambda^r({\cal A})$
(where $\Lambda^r({\cal A})$ are $\cal A$-bimodules and
$\Lambda^0({\cal A})= \cal A$) together with a linear operator
$d \, : \, \Lambda^r({\cal A}) \rightarrow \Lambda^{r+1}({\cal A})$
satisfying $ d^2=0$ and $ d(\omega \omega') = (d\omega) \, \omega'
+(-1)^r \omega \, d\omega' $ where $\omega \in \Lambda^r({\cal A})$.
We will assume that $\Lambda({\cal A})$ has a unit $\idty$ such that
$d \idty = 0$.
By now there is a vast literature dealing with differential calculi
on various types of mostly {\em non}-commutative algebras, in particular
quantum groups. But even {\em commutative} algebras exhibited in this
context rather unexpected features. In physics, models of elementary
particle physics were built with a space-time of the form $M \times
\Ir_2$ where $M$ is a four-dimensional differentiable manifold
\cite{Conn+Lott90}. Using differential calculus on (the algebra of
functions on) the two-point space $\Ir_2$, it was possible to extend the
Yang-Mills action to $M \times \Ir_2$. Its $\Ir_2$-part turned out to
be the usual Higgs potential. Later, it was demonstrated that a
certain deformation of the ordinary calculus of differential forms (on
$\Rl^n$) leads to lattice calculus \cite{DMH92,DMHS93}. In particular,
the Wilson action of lattice gauge theory was rederived as a
corresponding deformation of the usual continuum Yang-Mills action.
Also a relation with $q$-calculus \cite{Andr86} has been established
\cite{DMH92}. Some of these aspects will be briefly recapitulated in
section 2. In section 3 we discuss differential calculus on an
arbitrary discrete set (see also \cite{DMH93}). In particular, the
lattice calculus of section 2 is then recovered as a `reduction' of
the {\em universal} differential calculus (the `universal differential
envelope' of $\cal A$  \cite{Conn86,Coqu89}).
Relations with `posets' (partially ordered
sets) point towards some interesting applications in physics (cf
\cite{Sork91,BBET93}).

\section{Differential calculus and lattices}
\setcounter{equation}{0}
Let $x$ be the identity function on $\Rl$ and $\cal A$ the algebra of
$\Cx$-valued functions of $x$. The following commutation relation is
then consistent with the structure of a differential algebra,
\begin{eqnarray}          \label{f-dx}
     f(x) \, dx = dx \, f(x+a)  \qquad (\forall f \in {\cal A})
\end{eqnarray}
where $a \in \Rl, \, a >0$. The (formal) differential $dx$ is then
a basis of $\Lambda^1({\cal A})$ as a right (or left) $\cal A$-module.
This allows us to introduce right and left partial derivatives,
$\stackrel{\rightarrow}{\partial}_x$ and $\stackrel{\leftarrow}
{\partial}_x$ respectively, with respect to $x$ via
\begin{eqnarray}
   df = dx \, \stackrel{\rightarrow}{\partial}_x \! f
      = (\stackrel{\leftarrow}{\partial}_x \! f) \, dx  \; .
\end{eqnarray}
Then
\begin{eqnarray}
 dx \, \stackrel{\rightarrow}{\partial}_x \! f &=& {1 \over a} \,
    \lbrack x , dx \rbrack \, \stackrel{\rightarrow}{\partial}_x \! f
 = {1 \over a} \, \lbrack x , df \rbrack
 = {1 \over a} \, \left( f(x) \, dx - dx \, f(x) \right) \nonumber \\
 &=& dx \, {1 \over a} \, \left( f(x+a) - f(x) \right)
\end{eqnarray}
where we have used (\ref{f-dx}) and the Leibniz rule for $d$
shows that $\stackrel{\rightarrow}{\partial}_x$ is the {\em discrete}
(right) derivative. An {\em integral} associated with the differential
calculus should satisfy
\begin{eqnarray}                \label{int df}
  \int df = f + \mbox{`constant'}   \; .
\end{eqnarray}
In the case under consideration, a function $h$ should be called
`constant' if $dh=0$ which is the case iff $h$ is a periodic
function with period $a$. One finds that the condition (\ref{int df})
is sufficient to enable us to calculate $\int dx \, f(x)$ for any
$f \in \cal A$ \cite{DMH92}. Since the integral of a function is only
determined up to a periodic function,  a {\em definite} integral
is only well-defined over an interval the length of which is an integer
multiple of $a$. In this case,
\begin{eqnarray}
  \int_{x_o - m a}^{x_o + n a} dx \, f(x) =
  a \, \sum_{k=-m}^{n-1} f(x_o+k \, a) \; .
\end{eqnarray}
Hence, the integral (for $a>0$) is the usual approximation of the
Riemann integral.
\vskip.2cm

Let us consider a coordinate transformation $y = q^{x/a}$ where
$q \in \Cx$ is not a root of unity. For a function $f(y)$ we then have
\begin{eqnarray}
 dx \stackrel{\rightarrow}{\partial}_x \! f = df
 = dy \, \stackrel{\rightarrow}{\partial}_y \! f
 = dx \, \stackrel{\rightarrow}{\partial}_x \! y \,
         \stackrel{\rightarrow}{\partial}_y \! f
 = dx \, {q-1 \over a} \, y \, \stackrel{\rightarrow}{\partial}_y \! f
\end{eqnarray}
which leads to the $q$-{\em derivative}
\begin{eqnarray}
       \stackrel{\rightarrow}{\partial}_y \! f
      = { f(q \, y) - f(y) \over (q-1) \, y }   \; .
\end{eqnarray}
A similar expression is obtained for the left partial derivative
$\stackrel{\leftarrow}{\partial}_y$. The integral introduced above
yields in particular
\begin{eqnarray}
  \int_0^{\infty} dy \, f(y) = (q-1) \, \sum_{k=-\infty}^\infty
                               f(q^k) \, q^k  \; .
\end{eqnarray}
The rhs is defined in the mathematical literature as the
$q$-{\em integral} (see \cite{Andr86}, for example). It also makes sense
when $q^N =1, \, N \in \Ir$, which corresponds to the case of a closed
(i.e., periodic) lattice in $x$-space. Furthermore, we obtain
representations of the $q$-{\em plane} and $q$-deformed canonical
commutation relations, namely
\begin{eqnarray}
 \stackrel{\leftarrow}{\partial}_y \stackrel{\rightarrow}{\partial}_y
  - q \stackrel{\rightarrow}{\partial}_y
       \stackrel{\leftarrow}{\partial}_y \, = 0
 \quad , \quad
 \stackrel{\rightarrow}{\partial}_y  y - q \, y
         \stackrel{\rightarrow}{\partial}_y \, = 1
\end{eqnarray}
where $y$ has to be regarded as a multiplication operator.
These relations are known to be invariant under the coaction of the
quantum group $SL_q(2)$.
\vskip.2cm

There is an obvious generalization of the differential calculus to
higher dimensions \cite{DMHS93}. It provides us with a `universal'
framework for the deformation of continuum to lattice theories.
We refer to \cite{DMH92,DMHS93} for further details, results and
references.

\section{Differential calculus on a discrete set}
\setcounter{equation}{0}
Let us consider a discrete\footnote{Here by `discrete' we mean finite
or denumerable.} set $M$.
${\cal A}$ denotes the algebra of $\Cx$-valued functions on $M$
(with pointwise multiplication).
There is a distinguished set of functions $e_i$ on $M$, defined by
$e_i(j) = \delta_{ij}$ where $i,j \in M$. They satisfy the relations
\begin{eqnarray}
  e_i \, e_j = \delta_{ij} \, e_i   \qquad , \qquad
   \sum_i e_i= \idty
\end{eqnarray}
where $\idty$ denotes the constant function $\idty(i)=1$ which will
be identified with the unit in $\Lambda({\cal A})$.
Each $f\in {\cal A}$ can then be written as
\begin{eqnarray}
             f=\sum_i f(i) \, e_i   \; .
\end{eqnarray}
In the following we consider the {\em universal} differential
calculus on $\cal A$. From the properties of the set of functions $e_i$
we obtain
\begin{eqnarray}
        e_i \, de_j = - (de_i) \, e_j + \delta_{ij} \, de_i
        \quad , \quad
        \sum_i de_i = 0
\end{eqnarray}
(assuming that $d$ commutes with the sum)
which shows that the $de_i$ are linearly dependent. Let us introduce
\begin{eqnarray}
           e_{ij} := de_i \, e_j   \quad (i \neq j)
   \quad , \quad    e_{ii} := 0
\end{eqnarray}
(note that $de_i \, e_i \neq 0$) and
\begin{eqnarray}
 e_{i_1 \ldots i_r} := e_{i_1 i_2} e_{i_2 i_3} \cdots e_{i_{r-1}i_r}
  \qquad (r>1) \; .
\end{eqnarray}
Then
\begin{eqnarray}
 e_k \, e_{i_1 \ldots i_r} = \delta_{k i_1} \, e_{i_1 \ldots i_r}
     \quad , \quad
 e_{i_1 \ldots i_r} \, e_k = e_{i_1 \ldots i_r} \, \delta_{i_r k}
     \quad , \quad
 e_{i_1 \ldots i_r} \, e_{k \ell} = \delta_{i_r k} \,
 e_{i_1 \ldots i_r \ell}
\end{eqnarray}
and it can be shown that $e_{i_1\cdots i_r}$
with $i_k \neq i_{k+1}$ for $k=1,\ldots, r-1$ form a basis over $\Cx$
of $\Lambda^{r-1}({\cal A})$, the space of $(r-1)$-forms ($r>1$).
In particular, one finds
\begin{eqnarray}           \label{de_i}
   d e_i = \sum_j (e_{ij} - e_{ji})   \; .
\end{eqnarray}
\vskip.2cm

{}From the above relations it is obvious, that we can set some of the
$e_{ij}$ to zero without generating further relations or running into
conflict with the rules of differential calculus. This leads to
(consistent) {\em reductions} of the universal differential calculus.
They are conveniently represented by graphs in the following way.
We regard the elements of $M$ as vertices and associate with
$e_{ij} \neq 0$ an arrow from $j$ to $i$. The universal differential
calculus then corresponds to the graph where all the vertices
(corresponding to the elements of $M$) are connected pairwise by
arrows in both directions. Deleting arrows leads to graphs which
represent reductions of the universal calculus. Some interesting
examples of differential calculi arising in this way are considered in
the following.
\vskip.2cm
\noindent
(1) We choose $M = \Ir^n$ and impose the condition
\begin{eqnarray}
  e_{k \ell} \neq 0 \quad \Leftrightarrow \quad
  k = \ell + \hat{\mu} \mbox{ for some } \mu
\end{eqnarray}
where $\hat{\mu} := (0,\ldots,0,1,0,\ldots,0)$ with a 1 in the
$\mu$th entry. The associated graph is a lattice with an
orientation. Using
\begin{eqnarray}                    \label{x^mu}
   x^\mu := a \, \sum_k k^\mu \, e_k  \qquad (a \in \Rl, a >0)
\end{eqnarray}
we can express any function $f \in {\cal A}$ as $f(x)$.
The (reduced) differential calculus then implies
\begin{eqnarray}
      f(x) \, dx^\mu = dx^\mu \, f(x + \hat{\mu} \, a)
      \qquad (\mu = 1, \ldots, n)
\end{eqnarray}
which corresponds to the calculus considered in the previous section
and underlies the usual lattice theories (where $a$ is the lattice
constant).
\vskip.2cm
\noindent
(2) We consider the differential calculus on $\Ir^n$ associated
with the `symmetric lattice', i.e.,
\begin{eqnarray}
  e_{k \ell} \neq 0 \quad \Leftrightarrow \quad
  k = \ell + \hat{\mu} \mbox{ or }
  k = \ell - \hat{\mu}  \mbox{ for some } \mu  \; .
\end{eqnarray}
Let us define
\begin{eqnarray}
  e^{\pm \mu} := \sum_k e_{k \pm \hat{\mu}, k}  \quad , \quad
  \tau^\mu := a^2 \, ( e^{+\mu} + e^{-\mu} )  \; .
\end{eqnarray}
Using (\ref{de_i}), we obtain $dx^\mu = a \, ( e^{+\mu} - e^{-\mu} )$
with $x^\mu$ defined as in (\ref{x^mu}) and
\begin{eqnarray}
 \lbrack f(x) , dx^\mu \rbrack & = &
   {a^2 \over 2} \, dx^\mu \, \Delta_\mu f(x) + \tau^\mu \,
    \bar{\partial}_\mu f(x)           \\
 \lbrack f(x) , \tau^\mu \rbrack & = &
 a^2 \, dx^\mu \, \bar{\partial}_\mu f(x) + {a^2 \over 2} \, \tau^\mu
  \, \Delta_\mu f(x)
\end{eqnarray}
where
\begin{eqnarray}
 \partial_{\pm \mu} f & := & \pm {1 \over a} \, \left( f(x \pm a \,
                           \hat{\mu}) - f(x) \right)      \\
 \bar{\partial}_\mu f & := & {1\over2} \, (\partial_{+\mu} f +
                           \partial_{-\mu} f)             \\
 \Delta_\mu f & := & \partial_{+\mu} \, \partial_{-\mu} f =
 {1 \over a} \, (\partial_{+\mu} f - \partial_{-\mu} f)  \; .
\end{eqnarray}
Furthermore,
\begin{eqnarray}
 df & = & \sum_\mu (dx^\mu \, \bar{\partial}_\mu f + {1\over2} \,
        \tau^\mu \, \Delta_\mu f)  \; .
\end{eqnarray}
This differential calculus looks very much like a lattice version of a
(generalization of the) calculus which arises in the classical limit of
bicovariant differential calculus on the quantum group $SL_q(2,\Rl)$
\cite{MH+Reut93}.
\vskip.2cm
\noindent
(3) Let $S$ be a topological space (one may think of spacetime, for
example). In general, the collection of all open sets will be infinite.
A finite subtopology $\cal T$ may then be regarded as an approximation
of $S$ (related, e.g., to inaccurate position and time measurements).
Naturally associated with such a finite collection of open sets
covering $S$ is a graph in the following way. The vertices are
the elements of $\cal T$ (the open sets). If $U \subset V$ for
$U,V \in {\cal T}$ we draw an arrow from the vertex representing
$U$ to the one representing $V$. From the discussion above we know
that there is a differential calculus associated with such a
graph. These graphs (or `posets') generalize the ordinary lattice
and allow us to include nontrivial topology in discretized field
theories (see also \cite{BBET93}).

\end{document}